\documentstyle[prd,aps,preprint]{revtex}
\begin{document}

\title{A simple variational approach to the\\ quantum Frenkel-Kontorova
model}
\author{Choon-Lin Ho$^{1,2}$ and Chung-I Chou$^3$}
\address{\small \sl
$^1$Department of Physics, Tamkang University, Tamsui 25137,
Taiwan\footnote{Permanent address}\\
$^2$Theory Division, Institute of Particles and Nuclear Studies, KEK,\\
Tsukuba, Ibaraki 305, Japan\\
$^3$Institute of Physics, Academia Sinica, Taipei 11529,  Taiwan}


\maketitle

\begin{abstract}
We present a simple and complete variational approach to the one-dimensional
quantum Frenkel-Kontorova model.  Dirac's time-dependent variational principle
is adopted together with
a Hatree-type many-body trial wavefunction for the atoms.  The single-particle
state is assumed to have the Jackiw-Kerman form.  We obtain an effective
classical Hamiltonian
for the system which is simple enough
for a complete numerical solution for the static ground state of the model.
Numerical results show that our simple approach captures the essence of the
quantum effects first observed in quantum Monte Carlo studies.
\end{abstract}

\newpage

{\it 1. Introduction.}   The Frenkel-Kontorova (FK) model \cite{FK,FM} is a
simple
one-dimensional model used to study incommensurate structures appearing in
many condensed-matter systems, such as charge-density waves, magnetic spirals,
and adsorbed monolayers \cite{Bak}.  These modulated structures arise as a
result of the competition between two or more length scales.
The FK model describes a chain of atoms connected by harmonic springs
subjected to an
external sinusoidal potential.  In an important development in the study of
the classical FK model, Aubry \cite{Aubry} first made  use of the connection
between the
FK model,  the so-called ``standard map", and the Kolmogorov-Arnold-Moser
(KAM)
theorem to reveal many interesting features of the FK model.  Particularly, he
showed that when the mean distance (also called the winding number) between
two successive atoms is rational, the system is always pinned.  But when the
winding number is irrational, there exits a critical external field
strength below (above) which the system is unpinned (pinned).   This
transition
is called by Aubry a ``transition by breaking of analyticity", and is closely
connected with the breakup of a KAM torus.  It is very analogous to a phase
transition, and various critical exponents and questions of universality have
been extensively studied in the past.

In recent years, the FK model has been applied to the study of transmission in
Josephson junction and atomic-scale friction-nanoscale tribology \cite{JT}.
In these cases, quantum effects are very important.  Unlike the classical case,
study of quantum FK models is rather scanty.  It was first considered in a
quantum Monte Carlo (QMC) analysis in \cite{BGS}.  Their main observation
is that
the map appropriate to describe the quantum case is no longer the standard
map, but rather a map with a sawtooth shape.  An explanation of this
phenomenon was
later attempted in \cite{BBC} using a mean field theory in which the
inclusion of the
contribution from quasidegenerate states is essential.  But the mean field
computations with these states are rather involved, and the quantum map that
is to replace the classical standard map in different quantum regimes is not
clearly identified.
More recently, a less complicated approach was proposed in \cite{BLZ} which
uses the squeezed state
function to demonstrate that the sawtooth behavior is simply the result of
quantum fluctuations.  In our opinion, the approach adopted in \cite{BLZ} is
very appealing in principle.
However, we believe that some difficulties in this work need to be overcome
before it could be considered satisfactory.
First, the assumed squeezed state many-body ground state is general
enough so
as to include the correlations of the positions of the atoms, expressed by the
covariances $G_{ij}=\langle (x_i - {\bar x}_i)(x_j - {\bar x}_j)\rangle$
($i\neq j$),
where $x_i$ is the position of the $i$th atom, $\langle\cdots\rangle$ is the
expectation value in a given quantum state, and ${\bar x}_i=\langle
x_i\rangle$.  However, to find the equilibrium state of the model, one has to
solve
a system of coupled equations of the variables $x_i$ and the $G_{ij}$.  The
equations obtained are so complicated that the task of solving them within a
single
numerical framework is very difficult.  In fact, in \cite{BLZ} a hybrid
numerical
analysis was adopted in which the equations for the $G_{ij}$ were not solved.
Instead,
the values of $G_{ij}$ were taken from QMC data.  These values were then
treated as initial conditions in solving the equations for the atomic
positions
$x_i$'s.   Technically, such hybrid analysis is not satisfactory.  Second,
the covariance terms $G_{ij}$ ($i\neq j$) are constrained by the values
of the fluctuation terms $G_{ii}$ and $G_{jj}$ through the Cauchy-Schwarz
inequality.  These constraints guarantee the boundedness from below of
the effective Hamiltonian \cite{BL}.  But then this also calls for a proper
variational principle that has to take care of the interdependence of the
$G_{ij}$ terms.

In this letter we shall show that
all the essential features observed in the QMC studies can be obtained from
an independent-particle picture of the many-body ground state without the
covariance terms.  In the independent-particle picture the many-body trial
wavefunction are factorizable into single-particle states.  One can assume the
single-particle
state to have the form of a squeeze state.  For the quantum FK model,
a simpler and, in our
view, more elegant approach is to use
the Jackiw-Kerman (JK) function \cite{JK} as the single
particle state.  We shall show that this simple independent-particle approach
produces an effective
classical Hamiltonian which is bounded below, admits simple numerical
solution of the ground state without recourse to QMC analysis, and reproduces
the essential features observed in QMC studies.

{\it 2. Effective Hamiltonian.}   The Hamiltonian of the quantum FK model
is given by
\begin{equation}
{\cal H}=\sum_i \left[\frac{{\hat p}^2_i}{2m} +
\frac{\gamma}{2}\left({\hat q}_{i+1}-{\hat q}_i
\right)^2- V\cos(l_0{\hat q}_i) \right].
\label{FK1}
\end{equation}
Here ${\hat q}_i$ and ${\hat p}_i$ are the position and momentum operators,
respectively, of the $i$th atom, $\gamma$ the elastic constant of the
spring, $V$ and $2\pi/l_0$ are the strength and the period of the
external potential.  As in \cite{BBC}, it is convenient to use the
dimensionless variables
${\hat Q}_i=l_0{\hat q}_i$, ${\hat P}_i=l_0{\hat p}_i/\sqrt{m\gamma}$,
and $K=Vl_0^2/\gamma$.  With these new variables, we obtain
the following dimensionless Hamiltonian $H$
\begin{equation}
H=\sum_i \left[\frac{{\hat P}^2_i}{2} +
\frac{1}{2}\left({\hat Q}_{i+1}-{\hat Q}_i
\right)^2- K\cos({\hat Q}_i) \right].
\label{FK2}
\end{equation}
We have ${\cal H}=\gamma H/l_0^2$.  The effective Planck constant is
$\tilde\hbar=\hbar l_0^2/\sqrt{m\gamma}$.  For the classical FK model, the
Aubry transition occurs at the critical value $K_c=0.971635\cdots$.

To study the ground state properties of the quantum FK model in (\ref{FK2}),
we
adopt here the time-dependent variational principle pioneered by Dirac
\cite{Dirac}. In this approach, one first constructs the effective action
$\Gamma=\int dt
~\langle \Psi,t|i\hbar\partial_t -{\cal H}|\Psi,t\rangle$ for a given system
described
by $\cal H$ and $|\Psi,t\rangle$.  Variation of $\Gamma$ is then the  quantum
analogue of the Hamilton's principle.  The time-dependent Hatree-Fock
approximation emerges when a specific ansatz is made for the state
$|\Psi,t\rangle$.  We now assume the trial wavefunction of the ground state of
our quantum FK system to have the Hatree
form $|\Psi,t\rangle=\prod_i |\psi_i,t\rangle$, where the
normalized single-particle
state $|\psi_i,t\rangle$ is taken to be the JK wavefunction \cite{JK}:
\begin{eqnarray}
\langle Q_i|\psi_i,t\rangle&=&\frac{1}{(2\pi\tilde\hbar
G_i)^{1/4}}    
\times \exp\Biggl\{
-\frac{1}{2\tilde\hbar}\left(Q_i-x_i\right)^2\Bigl[\frac{1}{2}G_i^{-1}
-2i \Pi_i\Bigr]+\frac{i}{\tilde\hbar}p_i\left(Q_i-x_i\right)
\Biggr\}~.
\label{JK}
\end{eqnarray}
The real quantities $x_i(t)$, $p_i(t)$, $G_i(t)$ and $\Pi_i (t)$ are
variational parameters the variations of which at $t=\pm\infty$ are assumed to
vanish.  The JK wavefunction can be viewed as the $Q$-representation
of the squeeze state \cite{TF}.
We prefer to use the JK form since the physical meanings of the variational
parameters contained in the JK wavefunction are most transparent, as we
shall show below.  Furthermore, the JK form is in the general Gaussian form so
that integrations are most easily performed.

It is not hard to
check that $x_i$ and $p_i$ are the expectation values of the operators
${\hat Q}_i$ and ${\hat P}_i$: $x_i=\langle \Psi |{\hat Q}_i|\Psi\rangle$,
$p_i=\langle \Psi |{\hat P}_i|\Psi\rangle$.  Also, one has
$\langle \Psi |({\hat Q}_i-x_i)^2|\Psi\rangle=\tilde\hbar G_i$, and
$\langle \Psi |i\tilde\hbar\partial_t|\Psi\rangle=\sum_i
(p_i\dot{x}_i-\tilde\hbar
G_i\dot{\Pi}_i)$, where the dot represents derivative with respect to (w.r.t.)
time $t$. It is now clear that
$\tilde\hbar G_i$ is the mean fluctuation of the position of the $i$-th atom,
and that $G_i>0$.
With these expectation values, the (rescaled) effective action $\Gamma$ for
the dimensionless $H$ can be worked
out to be $\Gamma (x,p,G,\Pi)=\int dt ~[\sum_i
\omega_0^{-1}~(p_i\dot{x}_i+\tilde\hbar\Pi_i
\dot{G}_i)-H_{eff}]$, where $\omega_0=\sqrt{\gamma/m}$ is the angular
frequency
of the spring, and $H_{eff}=\langle\Psi|H|\Psi\rangle$ is the effective
Hamiltonian given by
\begin{eqnarray}
H_{eff}&=&\sum_i \frac{1}{2}\left[p_i^2+\tilde\hbar\left(\frac{1}{4}G_i^{-1}
+4\Pi_i^2 G_i\right)\right] \nonumber\\
&+& \sum_i\frac{1}{2}\left(x_{i+1}-x_i\right)^2 \nonumber\\
&+& \sum_i\frac{\tilde\hbar}{2}\left(G_{i+1}+G_i\right) \nonumber\\
&-& \sum_i K\exp\left(-\frac{\tilde\hbar}{2}G_i\right)\cos x_i~.
\label{Heff}
\end{eqnarray}
The last term in (\ref{Heff}) can be very easily  obtained  from
$\langle\Psi|F(Q_i)|\Psi\rangle=\sum_{m=0}^\infty F^{(2m)}(x_i)({\tilde\hbar
G_i})^m /(2m)!!$, where  $F^{(n)}(x)=\partial^n F(x)/\partial x^n$, and
$n!!\equiv n(n-2)(n-4)\cdots 1$.   Eq.~(\ref{Heff}) is
bounded from below.  One sees from the form of the effective action $\Gamma$
that $\Pi_i$ is the canonical
conjugate of $G_i$.

Varying $\Gamma$ w.r.t. $x,~p,~G$ and $\Pi$ then gives the
equations
of motion in the Hatree-Fock approximation.  Since we are mainly concerned
with
the static properties of the ground state of the quantum FK model, we must set
the time derivatives of these variables to zero.  This gives the equations
which determine the values of variational parameters corresponding
to
the equilibrium states (which include the ground state).  Equivalently, we can
obtain the equations for the equilibrium states by directly varying the
effective Hamiltonian $H_{eff}$ w.r.t. the variables.
Varying $H_{eff}$ w.r.t. $p_i$, $\Pi_i$, $x_i$ and $G_i$
give, respectively,
\begin{eqnarray}
p_i&=&0~,~~~~4\Pi_i G_i=0~,\label{e1}\\
x_{i+1}&-&2x_i +
x_{i-1}=K\exp\left(-\frac{\tilde\hbar}{2}G_i\right)\sin x_i~,\label{e2}\\
\frac{1}{4}G_i^{-2} &-&
K\exp\left(-\frac{\tilde\hbar}{2}G_i\right)\cos x_i -2=4\Pi_i^2 ~.
\label{e3}
\end{eqnarray}
The second equation in (\ref{e1}) implies $\Pi_i=0$ as $G_i>0$. This in turn
means that the right hand side of eq.(\ref{e3}) is equal
to zero:
\begin{equation}
\frac{1}{4}G_i^{-2} -
K\exp\left(-\frac{\tilde\hbar}{2}G_i\right)\cos(x_i) -2=0~.
\label{G}
\end{equation}
In the limit $\hbar=0$, eq.(\ref{e2}) is equivalent to the standard map.
We note that eq.(\ref{e2}) was also obtained in \cite{BLZ}.  This is because
in the formulation in \cite{BLZ} the
covariances $G_{ij}$'s decoupled from the $x_i$ and the fluctuations $G_{ii}$
($G_i$ in our case) in the variation of their Hamiltonian w.r.t $x_i$.
Unlike our case,
of course,  these covariance terms do actually influence the solutions of
(\ref{e2}) through other equations obtained by variation of the Hamiltonian
w.r.t. the $G_{ii}$ and $G_{ij}$.  And it is these equations that caused the
difficulties mentioned in the Introduction.  In particular, the values of the
$G_{ii}$ were input from the QMC data in order to solve for the $x_i$ in
(\ref{e2}).  Our simple approach, on the other hand, allows us to solve for
both the values of $x_i$ and $G_i$ coupled by eqs.(\ref{e2}) and (\ref{G})
consistently by a single numerical method.

>From $p_i=\Pi_i=0$ and (\ref{Heff}), we see that the problem of finding the
static ground state of the quantum FK model reduces to the problem of
minimizing  w.r.t. to $x_i$ and $G_i$ the following
effective potential
\begin{eqnarray}
V_{eff}&=& \sum_i\frac{1}{2}\left(x_{i+1}-x_i\right)^2 \nonumber\\
&+& \sum_i\frac{\tilde\hbar}{8}G_i^{-1}  +
\sum_i\frac{\tilde\hbar}{2}\left(G_{i+1}+G_i\right) \nonumber\\
&-& \sum_i K\exp\left(-\frac{\tilde\hbar}{2}G_i\right)\cos x_i~.
\label{Veff}
\end{eqnarray}
Eq.(\ref{e2}) and
(\ref{G}) are just the conditions $\partial V_{eff}/\partial x_i=0$ and
$\partial V_{eff}/\partial G_i=0$, respectively.

{\it 3. Numerical results.}   We numerically solve for the set of variables
$x_i$ and
$G_i$ which characterize the ground state using the Newton method
\cite{Schell}.
In all our numerical computations the winding number $P/Q=610/987$, which is
an approximation of the golden mean winding number $(\sqrt{5}-1)/2$,  is
used
with the periodic boundary condition $x_{i+Q}=x_i+2\pi P$. This winding number
is much more accurate than those used in previous works to approximate the
golden mean number, thus giving us better
accuracy in the computations of physical quantities related to the ground
state.  We emphasize
that all values of $x_i$ and $G_i$ are determined by the same numerical method
consistently.  In particular, we do not have to input the values of $G_i$ from
quantum Monte Carlo results in order to solve for the $x_i$.

Having obtained the values of $x_i$ which give the mean positions of the
quantum atoms in the chain, we can compare the results with the classical
configuration, following \cite{BGS}, in two ways: (1) by the quantum hull
function,
which is the plot of $x_i$ (mod $2\pi$) of the atoms against their unperturbed
positions $2\pi i P/Q$ (mod $2\pi$); (2) by the so-called $g$-function,
defined by
\begin{equation}
g_i\equiv K^{-1}~(x_{i+1} - 2 x_i + x_{i-1})
\label{gf}
\end{equation}
versus the actual atomic positions $x_i$.
>From (\ref{e2}), we also have
\begin{equation}
g_i=\exp\left(-\frac{\tilde\hbar}{2}G_i\right)\sin x_i~.
\label{gf2}
\end{equation}
Here $G_i$ is related to $x_i$ by eq.(\ref{G}).
We see from this equation that quantum fluctuations $G_i$ will modify the
shape
of the classical $sine$-map.  In addition to these two types of graphs, we
also plot the graph of $G_i$ against the unperturbed and the actual positions.
The formal graph was first introduced in \cite{BGS} to show the
strong correlation of the fluctuations of atoms' positions with their
unperturbed positions.  We introduce the latter type of graphs here since we
think that such graphs provide better picture about how the quantum
fluctuations of the atoms are related to their actual positions.

In Fig.~1 we show the four graphs mentioned above with different values of
$\tilde\hbar$ for the supercritical case $K=5$.  Fig.~1(a) shows the quantum
hull functions.  For small values of $\tilde\hbar$ the quantum hull function
consists of a countable set of steps discontinuities, just as in
the classical case: the atoms are in a
pinning phase.  In fact, the atoms are more likely to be located near the
valley of the external potential well, namely, near $x_i=0$ (mod $2\pi$).  As
the quantum effect increases, i.e., for increasing values
of $\tilde\hbar$, the quantum hull function gradually changes into a monotonic
analytic function, signifying that the system is entering the depinning phase.
There exists a critical value, approximately
$\tilde\hbar_c=6.58$ for $K=5$, above which the quantum hull function changes
from an nonanalytic function to an analytic one.
This is a quantum analogue of the Aubry transition in the classical case, and
can therefore be called the quantum Aubry transition.

Next in Fig.~1(b) we show the graphs of the $g$-function.  The curve defined
by
(\ref{gf2}) with $G_i$ satisfying (\ref{G}) are shown here as dashed curves
for different $\tilde\hbar$.
In the classical limit ($\tilde\hbar=0$) this curve is simply the
standard map ($sine$-curve).  As $\tilde\hbar$ increases, the amplitude of
the curve decreases.  For sufficiently large $\tilde\hbar$, the curve
resembles more closely a ``sawtooth" shape.  This is first noted in QMC study
in \cite{BGS}.  Here we see that it comes out very naturally from the equation
of motion (\ref{G}) and (\ref{gf2}).  In the supercritical
case ($K=5$), when $\tilde\hbar<\tilde\hbar_c$, the positions $x_i$ of the
atoms cover
only a subset of the $g$-curves.  This is in accord with the fact
that the atoms are in the pinning phase [{\it cf}. Fig.~1(a)].  As
$\tilde\hbar$ increases, the points
begin to spread along the $g$-curve.  When $\tilde\hbar>\tilde\hbar_c$, the
$g$-graph is completely covered as the system has entered the depinning phase.

Fig.~1(c) shows the quantum fluctuations $G_i$ plotted against the actual
atomic positions $x_i$.  The dashed curves represent the curves of
eq.(\ref{G})
for different $\tilde\hbar$. For small $\tilde\hbar$, the atoms are located
near $x_i=0$ (mod $2\pi$) with small values of $G_i$ which means, from
(\ref{JK}), that the wavefunctions are highly peaked at these positions. As
the
quantum effect increases, the external potential is so modified that now the
atoms could be found at other positions, but with atoms at $x_i=\pi$ (mod
$2\pi$) having
the largest value of $G_i$.   This indicates that wavefunctions of the atoms
near the
top of the potential are more extended with smaller amplitudes.  Again, when
$\tilde\hbar > \tilde\hbar_c$, the curves of (\ref{G}) are completely covered
by the solutions $x_i$.  To compare with the results in \cite{BGS}, we plot
the
values of $G_i$ against the unperturbed positions in Fig.~1(d).  One sees that
the values of $G_i$ are strongly
correlated
with the unperturbed positions, as first noted in \cite{BGS}.  For
$\tilde\hbar <\tilde\hbar_c$ the graphs
consists of steps discontinuities, and for $\tilde\hbar>\tilde\hbar_c$ the
graphs are continuous.  This is correlated with the graphs of the quantum
hull function in Fig.~1(a), since from (\ref{G}) any fixed value of $x_i$
correspond to a fixed value of $G_i$.

Next we show in Fig.~2 the corresponding graphs for the case $K=1.5$.
This represents the situation which is slightly over the critical classical
case.  The general trends of the behavior of the graphs are the same as
those in Fig.~1.  As expected, quantum Aubry transition takes place at a
smaller $\tilde\hbar_c=1.17$.  We note here that the shape of the
$g$-function at large $\tilde\hbar$ in this case is intermediate between a
$sine$ and a sawtooth map.

We have also checked the subcritical cases with $K<K_c$.  The
classical system is already in the depinning phase in this regime.  Quantum
fluctuations only enhance the trend of depinning.  The $g$-function is
found to be
closer to a $sine$-shape with smaller amplitude for higher $\tilde\hbar$.
This is consistent with the QMC results \cite{BGS}.

Finally, we note here that, while we have reproduced the essential features
first observed in the QMC studies of the quantum FK model, there is also
slight discrepancy between the
results of these two approaches.  The difference is that,
for a fixed value of $K$, the QMC results \cite{BGS} indicated that the
sawtooth shape of the $g$-function appeared at
a lower value of $\tilde\hbar$, and that the atoms began to spread along the
$g$-curve also at a smaller $\tilde\hbar$.  For example, at $K=5$ the
QMC
results showed that the above situation already appeared at $\tilde\hbar=0.2$,
while our results (cf. Fig.~1(b)) indicate that at $K=5$ and at a higher
$\tilde\hbar=2$ the system is still closer to the classical case.
We believe this could be explained as follows.  First, our
independent-particle
wavefunction is only the lowest order approximation of the many-body
wavefunction of the quantum FK system.  A more accurate description of the
system will require a better assumption of the wavefunction than that assumed
here.  This presumably may require the inclusion of the effects of the
covariance terms as advocated in \cite{BLZ}, but with a more appropriate
variational principle to circumvent the difficulties already mentioned in the
Introduction.  Second, our results are obtained at zero temperature, while
those in the QMC analysis were obtained, by the nature of the method itself, at
small but finite temperatures (temperature T=0.0067 as given in \cite{BGS}).
It
is natural that thermal fluctuations will cause the atoms to spread away from
their zero-temperature positions.

{\it 4. Summary.}   In conclusion, we have presented a simple and complete
variational approach
to the quantum FK model based on
a Hatree-type many-body trial wavefunction
of the JK form.
The effective Hamiltonian obtained is bounded below,
and is simple enough
for a complete numerical solution for the static ground state of the model
in various quantum regimes.
Numerical results show that our simple approach captures the essence of the
quantum effects first observed in QMC studies.  The map appropriate for the
quantum FK model is well described by eq.(\ref{gf2}) and (\ref{G}).
In contrast to previous approaches,  we do not require the existence of the
complicated quasidegenerate states, or the partial help from quantum Monte
Carlo data in order to obtain these results.

\vskip 1 truecm
\centerline{\bf Acknowledgments}

This work is supported in part by the
R.O.C Grant NSC 89-2112-M-032-004.  Part of the work was done while one of us
(CLH) was visiting the Theory Division at KEK (Japan) under the auspices of
the
exchange program between KEK (Japan) and the National Center for Theoretical
Sciences (Taiwan).  He would like to thank the
staff and members of the theory group of KEK for their hospitality and
support.  After the paper was submitted, we were kindly informed by Prof. B.
Hu that he and W.M. Zhang in their previous incomplete work \cite{Hu} had
obtained
a Hamiltonian similar to our eq.(\ref{Heff}), but that they had not studied
the ground state properties.

\vskip 3 truecm
\centerline{\bf Figures captions}
\begin{description}

\item[Fig.~1]  Structure of the quantum ground state for $K=5$ and winding
number $P/Q=610/987$ at $\tilde\hbar=2$ (black dots), $6$ (white dots) and $7$
(black curve). (a) quantum hull
function plotted against unperturbed atomic positions; (b) $g$-function
plotted
against actual atomic positions (the dashed curves represent eq.(\ref{gf2})
with $G_i$ satisfying (\ref{G}); (c) and (d) quantum fluctuations $G_i$
plotted against the actual and
unperturbed positions,  respectively.  The dashed curves in
(c) represent the curves of eq.(\ref{G}) for different $\tilde\hbar$.

\item[Fig.~2]  Same as Fig.~1 for  $K=1.5$ and $\tilde\hbar=0.5$
(black dots), $1.0$ (white dots) and $2$ (black curve).

\end{description}
\end{document}